\documentclass[12pt]{article}
\usepackage{epsf}
\usepackage{MnSymbol}
\newcommand{\be}{\begin{eqnarray}}
\newcommand{\ee}{\end{eqnarray}}
\newcommand{\ba}{\begin{array}}
\newcommand{\ea}{\end{array}}
\newcommand{\p}[1]{(\ref{#1})}
\newcommand{\lb}[1]{\label{#1}}

\def\bbox{{\,\lower0.9pt\vbox{\hrule \hbox{\vrule height 0.2 cm
\hskip 0.2 cm \vrule height 0.2 cm}\hrule}\,}}
\newcommand{\dsl}{\pa \kern-0.5em /}

\newcommand{\nn}{\nonumber \\}

\begin{document}

%%%%%%%%%%%%%%%% title page %%%%%%%%%%%%%%%%%%%%%%%%%%%%%%%%%%%%

\begin{titlepage}

\vfill

\begin{center}
\baselineskip=16pt {\Large Supercharges in the HKT Supersymmetric Sigma Models.}
\vskip 0.3cm
{\large {\sl }}
\vskip 10.mm
{\bf
A.~V. Smilga}
 \\

\vspace{6pt}
SUBATECH, Universit\'e de Nantes, \\
4 rue Alfred Kastler, BP 20722, Nantes 44307, France
\footnote {On leave of absence from ITEP, Moscow, Russia.}

\end{center}
\vspace{2cm}

\par
\begin{center} 
{\bf ABSTRACT}
\end{center}
\begin{quote}
We construct explicitly classical and quantum supercharges satisfying the standard ${\cal N} = 4$ 
supersymmetry algebra in the supersymmetric sigma models describing the motion over 
HKT (hyper-K\"ahler with torsion) manifolds. One member of the family of superalgebras thus obtained is equivalent to the superalgebra
derived and formulated earlier in purely mathematical framework.

\end{quote}
\end{titlepage}
%%%%%%%%%%%%%%%%%%%%%%%%%%%%%%%%%%%%%%%%

\setcounter{equation}{0}
\section{Introduction}

The HKT supersymmetric quantum mechanical (SQM) sigma models
\cite{HP} are known to enjoy the extended ${\cal N} = 4$ 
supersymmetry. 
\footnote{${\cal N}$ counts the number of real supercharges such that  the minimal supersymmetry (involving
double degeneracy of all excited states) corresponds to ${\cal N} = 2$.} 
That was derived earlier in the Lagrangian superfield framework \cite{GPS,Hull,DI}.
In this note we present explicit expressions for four real classical and quantum supercharges $Q^a$ and 
show that they satisfy the standard  ${\cal N} = 4$ algebra
 \be
\lb{N4alg}
 \{Q^a, Q^b\} \ = \delta^{ab} H \ .
 \ee

\section{Geometry}
To establish notations and bearing especially in mind a reader-physicist
\footnote{Unfortunately, mathematicians and physicists use  nowadays rather different languages, even when the problems they discuss are identical. In most of the cases, we do not understand each other without translation. This article is written in the mixture of two languages in a hope that it will be understandable to
both communities.} 
 we remind here basic definitions and properties of  
complex, K\"ahler, hyper-K\"ahler and HKT geometries.

\begin{itemize}

\item {\it  Complex manifold} is 
  a manifold of even dimension $D = 2d$ that can be
covered by several overlapping $D$-dimensional disks such that, in each disk, 
 complex coordinates $z^{j = 1,\ldots,d},
\bar z^{\bar j = 1,\ldots,d} $ can be chosen, with 
the metric having a Hermitian form
 \be
ds^2 \ =\ 2h_{j\bar k}(z, \bar z) dz^j d \bar z^{\bar k} \, , \ \ \ \ \ \ \ \
h_{j\bar k}^* = h_{k\bar j}\ .
 \ee
An important additional requirement is that, in the region where
a couple of such charts overlap, the  relationship between the coordinates in different charts is
 holomorphic, $\tilde z^j = f^j(z^k)$.

\item A necessary and sufficient condition for an even-dimensional manifold to be complex 
is the existence of the 
 tensor $I_{MN}$ (called  {\it  complex structure} tensor) satisfying the properties
 \be
\lb{Iprop1}
I_{MN} = - I_{NM}, \ \ \ \ \ \ \ I^{\ P}_{ N}  I^{\ M}_{P} = \ -\delta_N^M \ ,
 \ee

\be
\lb{huis}
\nabla_{[M} I_{N]\, P} \ = \ I^{\ Q}_{M}  I^{\ S}_{N}   \nabla_{[Q} I_{S]\, P} \ .
 \ee
The ``physical'' meaning of the second condition (vanishing of the so called Niejenhuis tensor) 
will be clarified below.

By a coordinate transformation, 
the tensor  $I^{\ N}_{M}$ acquires a simple canonical form
\be
\lb{Idiag}
 I \ =\ {\rm diag} (\epsilon, \ldots, \epsilon)\ , 
 \ee
where
\be
\lb{eps}
\epsilon = \ i\sigma^2 = 
\left( \begin{array}{cc} 0&1 \\ -1&0 \end{array} \right) \, ,
 \ee
In this frame, it is easy to introduce the complex coordinates: $z^1 = \frac 1{\sqrt{2}} (x^1 - ix^2)$, etc.
Under this choice, the complex components of the tensor  $I^{\ N}_{M}$ are   
 \be
\lb{Icompl}
I_m^{\ n} = - I^n_{\ m}   = -i \delta^n_m\, ; \ \ \ \ \ 
I_{\bar m}^{\ {\bar n}}= -  I^{\bar n}_{\ {\bar m}}    = i \delta^{\bar n}_{\bar m} \, .
  \ee 

\item A K\"ahler manifold is a complex manifold where the tensor $I_{MN}$ is covariantly constant,
 \be
\lb{nablaI}
\nabla_P I_{MN} = \partial_P I_{MN} - \Gamma^S_{PM} I_{SN} - \Gamma^S_{PN} I_{MS} =  0\ .
 \ee 
When bearing in mind \p{huis}, this requirement is equivalent to the requirement for the form
$K = I_{MN} dx^M \wedge dx^N$ to be closed, $dK  = 0$. 
 When $I_M^{\ N}$ is chosen as in \p{Icompl}, the condition \p{nablaI} implies that $\partial_j h_{l \bar k} -
\partial_l h_{j \bar k} = \partial_{\bar q} h_{j \bar k} - \partial_{\bar k} h_{j \bar q} = 0$.
 It can be represented as $h_{j \bar k} =  \partial_{j} \partial_{\bar k} K$. The function $K$ is called 
{\it K\"ahler potential}. 

\item For a generic complex manifold, $\nabla_P I_{MN} $ does not vanish for standard covariant derivatives
with symmetric Christoffel symbols. One can, however, consider  generalised covariant derivatives involving
torsions,
 \be
\hat{\Gamma}^M_{NK} = \Gamma^M_{NK} + \frac{1}{2}g^{ML}C_{LNK}\,, \lb{genGamma}
\ee
where $C_{LNK}$ is antisymmetric under $N \leftrightarrow K$. 
There are many such {\it affine connections} with respect to which $I_{MN}$ is covariantly constant. 
A particular such connection, the connection satisfying $  \tilde \nabla_P I_{MN} = 0$ 
with totally antisymmetric $C_{LNK}$ is called the {\it Bismut connection} \cite{Mavra,Bismut}. The explicit expression
for $C_{LNK}$ is \cite{Hull}
 \be
\label{Creal}
C_{LNK} = I_L^{\ P} I_N^{\ R} I_K^{\ T}  \left( \nabla_P I_{RT} +
\nabla_R I_{TP} + \nabla_T I_{PR} \right)\ .
  \ee
and one can observe that, for K\"ahler manifolds, this vanishes. 

\item A hyper-K\"ahler (HK) manifold is a manifold admitting three different covariantly constant
complex structures $I^{(1,2,3)} \equiv \{I,J,K\}$ satisfying the quaternionic algebra
 \be
\lb{Iab}
I^{(a)} I^{(b)} =  - \delta^{ab} + \epsilon^{abc} I^{(c)} \ .
 \ee 
A HK manifold has a dimension $D = 4n$. Locally, one can always choose coordinates where $ I^{(a)}$ acquire
a simple canonical form, 
 \be
\lb{canonI}
I = {\rm diag} ({\cal I}, \ldots, {\cal I})\, , \ \   
J = {\rm diag} ({\cal J}, \ldots, {\cal J})\, , \ \ 
K = {\rm diag} ({\cal K}, \ldots, {\cal K})
 \ee
with 
  \be
 \lb{Icanon}
 {\cal I} = \left( \begin{array}{cccc} 0&1&0&0 \\ -1&0&0&0 \\ 0&0&0&1 \\ 0&0&-1&0 \end{array} \right), \ 
{\cal J} = \left( \begin{array}{cccc} 0&0&0&1 \\ 0&0&1&0 \\ 0&-1&0&0 \\ -1&0&0&0 \end{array} \right) , \ 
{\cal K} = \left( \begin{array}{cccc} 0&0&1&0 \\ 0&0&0&-1 \\ -1&0&0&0 \\ 0&1&0&0 \end{array} \right)
 \ee
Note that these matrices are self-dual. 

\item Finally, a HKT manifold is, again, a $4n$-dimensional manifold with three quaternionic 
complex structures, with the
latter being covariantly constant  with respect to   the Bismut connection 
(required to be {\it one and the same} for all three complex structures)
rather than to the standard Levy-Civita connection.

It follows that the standard covariant derivatives of the complex structures are  
\be
\lb{nablHKT}
\nabla_P I^{(a)}_{MN} = \ \frac 12 g^{ST} (C_{TNP} I^{(a)}_{SM} - C_{TMP} I^{(a)}_{SN} ) 
 \ee
 with the same universal $C$. 

The simplest example of a HKT manifold is a conformally flat 4-dimensional manifold with the metric
 \be
\lb{metr4conf}
ds^2 \ =\ \frac {(dx^M)^2}{f^2} \, .
 \ee

%If 

 If we want  the metric  to stay nonsingular, the manifold can be made compact, if choosing  
 the  metric
  \be
ds^2 \ =\ \frac {d\bar z_j dz_j}{\bar z_k z_k}
 \ee
where the complex coordinates $z_{j=1,2}$ lie in the region $1 \leq |z_j| \leq 2$, with 
identification $z_j \equiv 2z_j$ when $|z| =1$.  This is the so called Hopf manifold. Topologically, it is $S^3 \times S^1$ or $SU(2) \times U(1)$.

The choice $f=1 + (x^M)^2/2$, also  describes a compact manifold ($S^4$), though
  such metric is singular at $x^M = \infty$. The latter may in principle lead to problems 
in defining supersymmetric Hilbert space \cite{flux}, but probably does not  in this case \cite{S4}.

\end{itemize}

\section{Supersymmetry}
Mathematicians consider different {\it complexes} associated with different kinds of manifolds above. 
It is known since \cite{Witten} that all of them can be interpreted in the framework of supersymmetric quantum
mechanics. For example, for any manifold, one can define the supersymmetric sigma model with the complex supercharges
\cite{Fred}
\be
\lb{QN2}
Q = \psi^M \left( \Pi_M -i \Omega_{M,AB}  \bar \psi^A \psi^B \right),  \ \ \ 
\bar Q = \bar \psi^M \left( \Pi_M -i \Omega_{M,AB}   \psi^A \bar \psi^B \right) \ .
\ee
Here $\Pi_M = -i \partial/\partial x^M$ is the canonical momentum, 
$\Omega_{M,AB}$ are standard spin connections,  
\be
\lb{Omega}
\Omega_{M,AB}
\ =\ e_{AN} ( \partial_M e^N_B + \Gamma^N_{MK} e^K_B)\ , 
 \ee
$\psi^M$ are complex Grassmann variables,
$\bar \psi^N = g^{NM} \partial/\partial \psi^M$, and $\psi^A = e_M^A \psi^M$.

 The nilpotent supercharges $Q, \bar Q$ realize the 
exterior derivative $d$ and its conjugate $d^\dagger$ of the de Rham complex. The Hamiltonian 
 \be
\lb{HamN2}
H = \frac 12 \{\bar Q, Q\} = \nonumber \\
 \frac 12 g^{MN} \left( \Pi_M -i \Omega_{M,AB}  \bar \psi^A \psi^B \right) 
\left( \Pi_N -i \Omega_{N,CD}  \bar \psi^C \psi^D \right) - \frac 12 R_{MNPQ} \bar \psi^M \psi^N \bar \psi^P \psi^Q
 \ee
is mapped to the covariant Laplacian acting on the forms.

For K\"ahler manifolds, the Hamiltonian \p{HamN2} admits an extra pair of supercharges forming (together with
\p{QN2} ) the extended ${\cal N} =4$ superalgebra. For hyper-K\"ahler manifolds, there are three such extra 
pairs giving ${\cal N} =8$ supersymmetry. 

The model \p{QN2}, \p{HamN2} involves a complex fermionic variable for each real coordinate. Another class of 
SQM models involve half as much fermionic degrees of freedom, a real fermionic operator
 $\psi^M \equiv \bar \psi^M$ 
for each coordinate $x^M$. $\psi^M$ obey the Clifford algebra $\{\psi^M, \psi^N\} = g^{MN}$ and can be mapped into
gamma matrices.  In \cite{IvSm}, such  models describing twisted Dolbeault complexes on a generic 
complex manifold were considered.  Twisting means, for mathematicians, 
the presence of an extra line bundle  and, for physicists, the presence of an extra Abelian gauge field. 
\footnote{One can as well consider the systems involving a non-Abelian field.}
The models involve two real supercharges. 
For a certain particular choice of the
gauge field ( see Eqs. \p{parW}, \p{barparW} below),  one of these supercharges has the form
\cite{Braden,Mavra}
 \be
\label{Qreal}
{\cal Q}  \ =\ \psi^M  \left[ \Pi_M -  \frac i2\,  \Omega_{M, BC} \psi^B \psi^C \right ] +
\frac i{12}\, C_{KLM}  \psi^K \psi^L \psi^M \ ,
 \ee
 with $C_{KLM}$ given in \p{Creal}. 
Bearing in mind the mapping $\psi^M \to \gamma^M/\sqrt{2}$, \p{Qreal} can be interpreted
as the Dirac operator with extra torsions of some particular form. For K\"ahler manifolds, the latter are absent. 

Another real supercharge is obtained from \p{Qreal} by commuting
it with the operator $F = \frac i2 I_{MN} \psi^M \psi^N$. 
When complex coordinates are chosen such that
the complex structure tensor is reduced to \p{Icompl}, the operator $F$ acquires the form 
$F = \frac 12[\psi^A, \bar \psi^A]$ ($A$ being the tangent space indices) and is interpreted as the fermion charge.

Here and in the following, it is more convenient technically to deal not with the quantum operators 
and  their (anti)commutators, 
but rather with their Weyl symbols, functions of the bosonic phase space variables
$\Pi_M, x^M$ and Grassmann variables $\psi^M$, and calculate their Poisson brackets with
 \be
\lb{PBdef}
\{P_M, x^N\}_{P.B.} = \delta_M^N, \, \ \ \ \ \  \{\psi^M, \psi^N\}_{P.B.} = ig^{MN}\,.
 \ee
 To be  precise, the 
quantum (anti)commutator of certain operators 
corresponds not to the Poisson bracket, but to the Gr\"onewold-Moyal bracket \cite{Moyal} of their Weyl symbols.
 Generically, GM brackets involve extra terms. 
However, for the commutators considered in this paper, these extra terms vanish. 

    To be still more precise, we need the Weyl symbol not of the covariant quantum supercharge 
\p{Qreal} acting on the Hilbert space equiped with the covariant measure
 \be
\lb{measure}
\mu = \sqrt{g}\, d^Dx \, ,
 \ee
 but of the operator $g^{1/4} {\cal Q} g^{-1/4} $ obtained from ${\cal Q}$ by a similarity
transformation and acting on the Hilbert space with the flat functional measure $d^Dx$ 
(see Ref.\cite{howto} for detailed discussions and explanations).  
One can show that this Weyl symbol  is given by the same expression \p{Qreal} as the quantum operator 
without any extra terms.
The Poisson bracket $\{{\cal Q}^{cl}, F^{cl}\}_{\rm P.B.}$ is
  \be
\lb{S}
 S^{cl} = \{{\cal Q}^{cl}, F^{cl}\}_{\rm P.B.} = 
\  \psi^N I^{\ M}_{N} \left[ \Pi_M -   \frac i2\,  \Omega_{M, BC} \psi^B \psi^C  - 
\frac i4 C_{MKL} \psi^K \psi^L \right] \ .
 \ee
To derive it, it is convenient to represent again $\Pi_M$ as the operator $-i\partial_M$ and
notice that the structure 
$\partial_M +   \frac 12\,  \Omega_{M, BC} \psi^B \psi^C $ is nothing but the spinor covariant derivative. 
Then, profiting from the scalar nature of $F$, we can upgrade it to the full  covariant derivative acting also
  on the tensor indices like in \p{nablaI}, and use finally
the identity $\nabla_M \psi^N = 0$ (where $\nabla_M$ involves {\it both} the Christoffel and spinor parts).
Indeed, the spin connection \p{Omega}
is {\it defined} such that its contribution in $\nabla_M \gamma^N  \equiv \sqrt{2}  
\nabla_M \psi^N $ cancels other contributions.

 Ordering this with the  Weyl symmetric prescription and performing
the inverse similarity transformation to obtain the operator acting on the same
  Hilbert space as \p{Qreal}, we derive a beautiful result: similarly to the case of ${\cal Q}$, 
the quantum supercharge $S$ keeps the  form 
\p{S} with the operator order prescribed there.

Note now that the operators \p{Qreal}, \p{S} satisfy the minimal supersymmetry 
algebra, ${\cal Q}^2 = S^2 \equiv H, \ \{{\cal Q}, S \} = 0$,
if and only if the condition \p{huis} is satisfied. This {\it is} the physical meaning 
of this condition, it is necessary
for supersymmetry to hold. 

For mathematicians, the condition \p{huis} is necessary to define the Dolbeault complex. 
Indeed, the combinations 
$Q = {\cal Q} + iS$ and $\bar Q = {\cal Q} -iS$ can be mapped into the 
holomorphic exterior derivative $\tilde\partial$ and its conjugate 
$\tilde \partial^\dagger$. The notation $\tilde\partial$ means the presence of an extra 
twisting,
\be
\lb{parW}
\tilde\partial = \partial + A = \frac 12 \partial_M  \left( \delta^M_N + i I^{\ M}_{N} \right)dx^N + 
  \frac 1{16} \left( \partial_M \ln \det g \right) \left( \delta^M_N + i I^{\ M}_{N} \right)\, dx^N  \, . 
 \ee 
In other words, the Dirac complex is equivalent to the Dolbeault complex with some particular 
Abelian 
gauge field (in the mathematical language, such $A_M$ is the connection of the square root $K^{1/2}$ of 
the canonical line bundle).

The mapping Dirac $\leftrightarrow$ Dolbeault  means also the mapping of Hilbert spaces. In the Dirac interpretation, 
the quantum supercharges are expressed via $\gamma$-matrices and act upon the spinor wave functions. In the Dolbeault
interpretation, wave functions depend on the coordinates $x^M$ and the {\it holomorphic} fermion variables $\psi^m$.
\footnote{To avoid confusion, please, note that the derivative operator in \p{parW} acts on the coefficients $A, A_m, A_{mn}$, etc. 
of the expansion of such a wave function over $\psi^m$, but not on the variables $\psi^m$. This is in contrast to the expressions like
\p{Qreal}, where the operator $\Pi_M = -i\partial_M$ acts also on $\psi^N = e^{AN}(x) \psi^A$. }
 In the mathematical language, the Hilbert space consists of holomorphic $(p,0)$ forms and is denoted $\Lambda^{(p,0)}$. It is thus smaller than the Hilbert
space of the de Rahm complex $\Lambda^{(p,q)}$  involving {\it all} forms.

By the same token, the pure Dirac complex can be mapped to the anti-Dolbeault twisted complex with
 \be
\lb{barparW}
\tilde {\bar\partial} = \bar\partial + 
\frac 1{16} \left( \partial_M \ln \det g \right) \left( \delta^M_N - i I^{\ M}_{N} \right)\, dx^N  \, . 
 \ee 
The Hilbert space consists then of antiholomorphic $(0,q)$ - forms. 
 
The mappings  Dirac $\leftrightarrow$ Dolbeault  and Dirac $\leftrightarrow$ anti-Dolbeault are well known to mathematicians 
 \cite{Nicola}. A physicist may consult Ref. \cite{IvSm} for
further pedagogical explanations.
  
In this paper, we are discussing only SQM systems, 
 where the notion of chirality does not exist. However, each such  SQM sigma model 
can be upgraded to a certain  2-dimensional field theory  where supercharges are attributed with chirality. One can talk then about
the $(m,n)$ models with $m$ chiral and $n$ antichiral supercharges. Bearing this in mind, a generic Hamiltonian \p{HamN2} is 
$(1,1)$ -- supersymmetric,
the Hamiltonian \p{HamN2} for a K\"ahler Hamiltonian is $(2,2)$ -- supersymmetric, and the model with the supercharges 
\p{Qreal}, \p{S} can be though of as 
 $(2,0)$ -- supersymmetric or $(0,2)$ -- supersymmetric depending on whether it is associated it with the twisted Dolbeault or with the
twisted anti-Dolbeault complex. 

\section{HK and HKT.}

\vspace{.2cm}

\subsection{Hyper-K\"ahler manifolds.}

\vspace{.2cm}

Hyper-K\"ahler manifolds involve three different complex structures and, correspondingly, three different 
fermion charges
$F^{(a)} =   \frac i2 I_{MN}^{(a)} \psi^M \psi^N$. 
Bearing this in mind, one immediately constructs four real supercharges with Weyl symbols  
\be
\lb{QHK}
{\cal Q} =  \psi^M  \left[ \Pi_M - \frac i2\,  \Omega_{M, BC} \psi^B \psi^C \right]\, ,  \nonumber \\
S^{(1,2,3)} = \{{\cal Q}, F^{(1,2,3)}\}_{P.B.} = 
  \psi^N \left( I^{(1,2,3)}\right)^{\ M}_{N} \left[ \Pi_M   - \frac i2\,  \Omega_{M, BC} \psi^B \psi^C 
 \right]\, . 
\ee

A hyper-K\"ahler manifold is K\"ahler with respect to each  complex structure. It immediately follows
that 
 \be 
 \lb{subset}
\{{\cal Q}, {\cal Q} \}_{P.B.} = \{S^{(a)}, S^{(a)} \}_{P.B.} = 2iH\, , \ \ \ \ \ \ \ \  
\{{\cal Q}, S^{(1,2,3)} \}_{\rm P.B.} =0\, .
 \ee
 To find the bracket
 $\{S^{(a)}, S^{(b)} \}_{\rm P.B.} $ when $a \neq b$, 
consider first the bracket $\{S^{(a)}, F^{(b)} \}_{\rm P.B.}$. 
It is calculated using the same trick as was used when calculating the bracket 
$\{{\cal Q}, F \}_{\rm P.B.}$ above
(see the footnote before Eq.\p{S}).
 It is not difficult to derive  \cite{Fig}
 \be
\lb{SFHK}
 \{S^{(1)}, F^{(2)} \}_{\rm P.B.} \ =\ - S^{(3)},\ \  {\rm and \ cyclic \ permutations} .
  \ee  
Using now the Jacobi identity 
\be
\lb{Jacobi}
\{S^{(1)}, \{F^{(2)},{\cal Q} \}_{\rm P.B.} \}_{\rm P.B.} - \{F^{(2)}, \{ {\cal Q}, S^{(1)} \}_{\rm P.B.} \}_{\rm P.B.} - \{ {\cal Q}, \{  S^{(1)}, F^{(2)}\}_{\rm P.B.} \}_{\rm P.B.} = 0 
 \ee
(with minuses taking account of the odd nature of ${\cal Q}$ and $S^{(1)}$ and the even nature of $F^{(2)}$), 
it is straightforward to see that the bracket $ \{ S^{(1)},S^{(2)} \}$ 
as well as the brackets  $ \{ S^{(1)},S^{(3)} \}_{P.B.}$
and  $ \{ S^{(2)},S^{(3)} \}_{P.B.}$  
vanish, giving together with \p{subset}
the ${\cal N} = 4$ supersymmetry algebra \cite{Wipf}.

It was futher noticed in \cite{Wipf} that ${\cal N} = 4$ supersymmetry  
is also kept for a generalized system obtained from \p{QHK} by adding 
the gauge field, $\Pi_M \to \Pi_M - A_M$, provided the field strength tensor 
commutes with all complex structures, 
 \be
\lb{[FI]}
F_{MN} \left( I^{(a)} \right)^N_{\ P} =  \left( I^{(a)} \right)_M^{\ \ N}F_{NP} \,.
 \ee
 The field may be Abelian or non-Abelian. 
There is an additional requirement for the topological charge to be integer. 
 For a mathematician, this means that the field is a connection of a well-defined fiber bundle. For a physicist, 
it is necessary to keep the Hilbert space of the quantum system supersymmetric \cite{flux}. 
The conditions \p{[FI]} imply that, under the canonical frame choice  where the complex structures have the 
form \p{canonI}, \p{Icanon},    the tensor $F_{MN}$ is  anti-self-dual. To prove it, note that any antisymmetric
$4 \times 4$ matrix  can be represented as a linear combination of three self-dual matrices \p{Icanon} and 3 anti-self-dual
matrices
  \be
 \lb{Itilde}
 \tilde {\cal I} = \left( \begin{array}{cccc} 0&1&0&0 \\ -1&0&0&0 \\ 0&0&0&-1 \\ 0&0&1&0 \end{array} \right), \ 
\tilde {\cal J} = \left( \begin{array}{cccc} 0&0&0&1 \\ 0&0&-1&0 \\ 0&1&0&0 \\ -1&0&0&0 \end{array} \right) , \ 
\tilde {\cal K} = \left( \begin{array}{cccc} 0&0&1&0 \\ 0&0&0&1 \\ -1&0&0&0 \\ 0&-1&0&0 \end{array} \right)
 \ee
All three matrices in \p{Itilde} commute with ${\cal I},{\cal J}$ and ${\cal K}$. Using the commutation relations \p{Iab}, it is easy to see
now that, for any matrix $F = a_1 {\cal I} + a_2 {\cal J} + a_3 {\cal K} + b_1 \tilde {\cal I} + b_2 \tilde {\cal J} + b_3 \tilde {\cal K}$ commuting with ${\cal I},{\cal J},{\cal K}$,
 the coefficients $a_j$ necessarily vanish.
\vspace{.2cm}

\subsection{HKT manifolds.}

\vspace{.2cm}

Consider the supercharge \p{Qreal} and three supercharges \p{S} for three available complex structures.
The properties \p{subset} follow immediately. To show that the brackets  like 
$ \{ S^{(1)},S^{(2)} \}_{P.B.}$ vanish we have to show 
that the property \p{SFHK} holds also in the HKT case.

Like in the HK case, we can upgrade the spinor covariant derivative in \p{Qreal} up to the full covariant derivative
and use the identity $\nabla_M \psi^N = 0$. The novelties, however, are:
{\it (i)} the presence of the extra term $\propto CI^{(a)} \psi^3$ in $S^{(a)}$; 
 {\it (ii)} the fact that the covariant derivatives of $I^{(a)}$ do not vanish  anymore, 
but are given by the expressions
\p{nablHKT}.

As a result, the bracket $ \{S^{(1)}, F^{(2)} \}_{\rm P.B.}$ seems to involve besides $-S^{(3)}$ an extra term,
 \be
\lb{extra}
X \propto I_M^{\ P}  J_N^{\ R} C_{PRQ} \psi^M \psi^N \psi^Q \, .
 \ee
A remarkable fact, however, is that $X$ vanishes identically. To see this, assume that the complex structures are
 reduced to their canonical block-diagonal form \p{canonI} and consider one of the $4 \times 4$ blocks. One can then
represent
 $$ C_{PRQ} = \epsilon_{PRQS} A^S, \ \ \ \ \ \ \ \psi^M \psi^N \psi^Q = \epsilon^{MNQT} \chi_T\ . $$
Convoluting the epsilon tensors, one obtains 6 terms. Four of them vanish right away due to antisymmetry of 
$I,J$. And two remaining terms cancel each other due to anticommutativity of two different complex structures.

The relation $X=0$ can be expressed in terms of a nice identity
 \be
  \lb{mystery}
\left( I_M^{\ P}  J_N^{\ R} - I_N^{\ P}  J_M^{\ R} \right) C_{PRQ} +  \left( I_N^{\ P}  J_Q^{\ R} - I_Q^{\ P}  J_N^{\ R} \right) C_{PRM}
 + \left( I_Q^{\ P}  J_M^{\ R} - I_M^{\ P}  J_Q^{\ R} \right) C_{PRN}  = \ 0\, ,
   \ee
 which holds for two different complex structures $I,J$ and any antisymmetric $C_{PRQ}$. 

It follows, bearing in mind \p{Jacobi}, that the supercharge \p{Qreal} and three supercharges \p{S} satisfy the 
standard ${\cal N} = 4$ superalgebra, the same as in the hyper-K\"ahler case.  

${\cal N}=4$ supersymmetry holds also for a generalized system involving an 
anti-self-dual gauge field (or a self-dual one in the frame with the opposite orientation where the complex structures have the form  \p{Itilde} 
\cite{IvLech,Maxim,DI}). It can be proven in exactly the same way as in the hyper-K\"ahler case, the presence of nonzero torsion tensor $C_{PRQ}$ being irrelevant.

\section{Mathematical interpretation.}

These results can be translated into the mathematical language. We start by reminding that

\begin{itemize}
\item ${\cal N} = 4$ supersymmetry (or, bearing in mind the remark at the end of section 3, ${\cal N} = (2,2)$ supersymmetry) 
for K\"ahler manifolds is realized by two pairs of mutually 
conjugate 
nilpotent
operators, $\partial ,\partial^\dagger$ and $\bar\partial, \bar \partial^\dagger$, 
where $\partial$ ($\bar \partial$) is the holomorphic (antiholomorphic) exterior derivative.

\item For a generic complex manifold, $\partial$ does not anticommute anymore with 
 $\bar\partial^\dagger$, and 
only a ${\cal N} = 2$ subalgebra of this  
${\cal N} = 4$ superalgebra survives. Three choices of such a  subalgebra are possible. The pair
$\partial, \partial^\dagger$ [associated with ${\cal N} = (2,0)$] realizes the Dolbeault complex, the pair $\bar\partial, \bar\partial^\dagger$
[${\cal N} = (0,2)$]
 --- anti-Dolbeault complex, and one can also choose the pair  $d, d^\dagger$ where 
$d = \partial + \bar\partial$ is the total exterior derivative,  
which realizes the de Rham complex [${\cal N} = (1,1)$]. One can also consider twisted de Rham and Dolbeault complexes. The most relevant
for us is the  simplest version
of the twisted Dolbeault complex that amounts to adding to $\partial$ an exact holomorphic form as in 
\p{parW}, \p{barparW}. 

\item Hyper-K\"ahler manifolds enjoy   ${\cal N} = 8$ (or ${\cal N} = (4,4)$ )   supersymmetry realized by 4 conjugate pairs\cite{Verb}
$d,d^\dagger$, and $d_{1,2,3}, d_{1,2,3}^\dagger$, where $d_a = \partial_a - \bar \partial_a$,  
$\partial_a$ ($\bar\partial_a$) being holomorphic (antiholomorphic) exterior derivatives associated with the 
complex structure $I^{(a)}$. The only nonzero anticommutators are
 \be
\lb{N8}
\{d, d^\dagger\}  = \Delta, \ \ \ \ \ \ \{d_a, d^\dagger_b \} = \delta_{ab} \Delta \, ,
 \ee   
where $\Delta$ is the Laplacian.

\item For HKT manifolds the algebra \p{N8} does not hold and ${\cal N} = 8$ supersymmetry
is broken. 
%Generically, down to ${\cal N} = 2$. 

\end{itemize}

The main observation of this paper is that it is still, however, 
possible to keep ${\cal N} = (4,0)$ supersymmetry,
if considering {\it twisted} exterior derivatives \p{parW}, \p{barparW}.

Indeed, our supercharges can be mapped to
 the set
 \be
\lb{QN4math}
 {\cal Q} = \partial_1 + \partial_1^\dagger = \partial_2 + \partial_2^\dagger = \partial_3 + \partial_3^\dagger ; 
\nonumber \\
 S_a \ =\ i(\partial_a - \partial_a^\dagger) \, 
 \ee
where each $\partial_a$ is given by \p{parW} that  
involves the projectors associated with the complex structure $I^{(a)}$ and the Hermitian conjugation refers to the ``large'' 
Hilbert space of the de Rham complex. 
Note that the first line in \p{QN4math} is an identity, which holds for such twisted exterior derivatives and 
their conjugates,
 but does not hold for usual derivatives - neither in the HKT, nor in the HK case.

Obviously, our supercharges can also be mapped to the ${\cal N} = (0,4)$ set
\be
\lb{barQN4math}
 \bar Q = \bar\partial_1 + \bar\partial_1^\dagger = \bar\partial_2 + \bar\partial_2^\dagger = 
\bar\partial_3 + \bar\partial_3^\dagger ; 
\nonumber \\
 \bar S_a \ =\ i(\bar\partial_a - \bar\partial_a^\dagger) \, .
 \ee
with  antiholomorphic derivatives \p{barparW}. 

Note now that one can twist the derivatives $\partial_a$ and $\partial_a^\dagger$ still further by replacing
$$\partial_a \to \partial_a -i \left[ \delta_N^M  + i\left( I^{(a)} \right)_{N}^{\ M} \right] A_M dx^N \ , $$ 
 where $A_M dx^M$ is a  bundle (not necessarily line bundle) satisfying the condition discussed above: $F_{MN} = \partial_{[M} A_{N]}$ should commute with all complex structures
meaning that it is anti-self-dual in the canonical frame \p{canonI}, \p{Icanon}.  Such a deformation leaves supersymmetry intact.

Let us  establish  the correspondence between our findings and the results of Ref.\cite{Verb} where the presence of  an
 ${\cal N} = 4$ superalgebra  for HKT manifolds was demonstrated in the purely mathematical framework. 
This algebra  (defined at the end of Sect. 10.1 there)
  involves the operators acting on the Hilbert
space of the Dolbeault complex associated with one 
of the complex structures (say, $I$). The odd generators include the exterior {\it twisted} 
holomorphic derivative 
$\tilde\partial = \partial + \theta \equiv \partial_1+ \theta/2$ ($\theta$ being the connection of the canonical line bundle $K$),  its conjugate  $\tilde\partial^\dagger$, the 
operator 
 \be
\lb{deltaJ}
\tilde\partial_J = - J \circ \left(\bar \partial + \frac {\bar\theta}2 \right)  \circ J \, , 
 \ee
 and its conjugate.  
\footnote{In section 7 of this paper, the superalgebra involving untwisted $\partial$ and $\partial_J$ was discussed, but a commutator like $\{\partial^\dagger, \partial_J\}$ vanishes only  for a metric with constant determinant $g$, if the Hermitian conjugation is defined in a standard way with the covariant measure 
\p{measure} .}
 The explicit form of the first term in the R.H.S. of Eq.\p{deltaJ}
(the operator of antiholomorphic with respect to $I$ derivative conjugated by $J$)
can be derived as follows:
 \begin{itemize}
\item Choose the canonical frame and express the operator $\bar \partial$ via the complex coordinates $w^j, \bar w^{\bar j}$ 
corresponding to the complex structure
$J$. 
 \item  Change the sign of the terms involving $d \bar w^{\bar j}$.
 \item 
Reexpress everything in terms of the original variables $z^j, \bar z^{\bar j}$ associated with $I$.
 \end{itemize}

One obtains as a result 
 \be
\lb{deltaJJ}
 \partial_J \ =\ \sum_{4\times 4\  {\rm blocks}} \left( dz^2 \frac \partial {\partial \bar z^{\bar 1}} -   
dz^1 \frac \partial {\partial \bar z^{\bar 2}}\right) \,.
\ee
Note that both $\partial$ and $\partial_J$ are $SU(2)$ singlets. Note also that, if replacing $J$ by $K$ in \p{deltaJ}, one obtains
the {\it same} operator up to a factor $i$.
 
Let us compare now this with \p{barQN4math}. Consider only holomorphic with respect to $I$ forms.
As was explained above, the supercharges \p{Qreal}, \p{S} are  then mapped into the operators 
$\tilde \partial \pm  \tilde \partial^\dagger$
with twisted $\tilde \partial = \partial + \frac 14 (\partial_j \ln \det h ) dz^j$.

 Consider now the operator 
 \be
\lb{S+}
S_+ \ =\ -\frac 12 \psi^N (J + iK)^{\ M}_{N} \left[ \Pi_M -   \frac i2\,  \Omega_{M, BC} \psi^B \psi^C  - 
\frac i4 C_{MKL} \psi^K \psi^L \right]\ .
 \ee
Choosing the canonical frame with \p{canonI} and \p{Icanon}, it is not difficult to show that $S_+$ is mapped into
 \be
\lb{S+map}
S_+ \ \to \sum_{4\times 4 \ {\rm blocks}} ( dz^2 \tilde\partial_{\bar 1} - dz^1 \tilde\partial_{\bar 2} )
 \ee
with 
\be
\lb{obkladki}
\tilde\partial_{\bar j} = \partial_{\bar j} + \frac 14 (\partial_{\bar j} \ln \det h ) = \ (\det h)^{-1/4} \partial_{\bar j} (\det h)^{1/4}\, .
 \ee

We see that the supercharge  \p{S+} is related to the supercharge \p{deltaJJ} in exactly the same way as $\tilde \partial$ 
to $\partial$ : both involve the ``dressed'' derivative operators \p{obkladki}.

It is instructive to see what happens for  the simplest type of HKT manifolds with the metric \p{metr4conf}. The supercharges
for this model were found in \cite{Maxim}. Translating them to mathematical notation, they acquire the form 
that coincides with
\p{barQN4math}, 
   \be
\lb{4delty}
 i\tilde \partial &=& if \partial_j \frac 1f  dz_j \wedge  \, , \nn
-i\tilde \partial^\dagger &=& i f \bar\partial_{j} \frac 1f dz_j \righthalfcup\, , \nn
S_+ &=& i\epsilon_{jk}\, f \bar \partial_k \frac 1f dz_j \wedge \, , \nn
 S_- = S_+^\dagger &=&   i\epsilon_{jk}\, f \partial_k \frac 1f dz_j \righthalfcup \ ,
 \ee
where $\wedge$ stands for the exterior and $\righthalfcup$ for the interior product, $dz_j \righthalfcup dz_k = f^2 \delta_{jk}$,
and we do not distinguish here between covariant and contravariant indices. 
Actually, $dz_j \wedge$ and $dz_j \righthalfcup$ are nothing but $\psi_j$ and $\bar\psi_j$ in disguise, 
but one has to bear in mind that the derivatives
do not act here on the fermion operators. Bearing in mind the covariant norm \p{measure}, the presence of the ``dressings'' 
$f \cdots 1/f$ is essential for the first and  the second pairs of the operators in \p{4delty} to be mutually conjugate. 

As was mentioned above, the operators $\tilde \partial$, $\tilde \partial_J$ can be further twisted by adding
the connection with antiselfdual curvature. This defines a family of superalgebras of which Verbitsky's
one represents a particular member.

 \section{Acknowledgements}

I am indebted  to E. Ivanov for useful discussions (especially of the paper \cite{DI})  and to 
M. Verbitsky for many illuminating discussions and comments.

\end{document}